\documentstyle{article}
\begin{document}
\large
\newcommand{\mincir}{\ \raise -2.truept\hbox{\rlap{\hbox{$\sim$}}\raise5.truept
	\hbox{$<$}\ }}
\newcommand{\magcir}{\ \raise -2.truept\hbox{\rlap{\hbox{$\sim$}}\raise5.truept
 	\hbox{$>$}\ }}
\newcommand{\refer}{\par\noindent\hangindent 20pt}
\baselineskip=24pt

\bigskip
\bigskip
\begin{center}
{\bf THE GUNN-PETERSON EFFECT IN THE SPECTRUM OF THE Z=4.7
QSO 1202-0725: THE INTERGALACTIC MEDIUM AT VERY HIGH REDSHIFTS$^{\textstyle
\star}$} \end{center}

\bigskip
\noindent
E. Giallongo$^{1,2}$, S. D'Odorico$^2$, A. Fontana$^3$, R. G. McMahon$^4$, S.
Savaglio$^5$, S. Cristiani$^6$, P.Molaro$^7$, D. Trevese$^8$

\vfill
\noindent
$^1$~Osservatorio Astronomico di Roma, I-00040 Monteporzio, Italy\\
$^2$~European Southern Observatory, Karl Schwarzschild Stra{\ss}e, 1, 85748
Garching bei M\"unchen, Germany\\
$^3$~Dipartimento di Fisica, II Universit\`a di Roma, via E. Carnevale, I-00173
Roma, Italy\\
$^4$~Institute of Astronomy, Madingley Road, Cambridge CB3 0HA, UK\\
$^5$~Dipartimento di Fisica, Universit\`a della Calabria, I-87036 Arcavata
di Rende, Cosenza, Italy\\
$^6$~Dipartimento di Astronomia, Universit\`a di Padova, vicolo
dell'Osservatorio 5, I-35122 Padova, Italy\\
$^7$~Osservatorio Astronomico di Trieste, via G. B. Tiepolo 11,
I-34131 Trieste, Italy\\
$^8$~Istituto Astronomico, Universit\`a di Roma, ``La Sapienza'', via
G. M. Lancisi 29, I-00161 Roma, Italy\\

\vfill
$^{\textstyle \star}$Based on material collected at the ESO-La Silla
telescopes.

\newpage
\centerline{\bf ABSTRACT}
\bigskip

A measure of the average depression between Lyman absorption lines in
the spectrum of the faint quasar BR1202-0725 ($z_{em}=4.695$) is
presented.
The relatively high resolution of the spectrum ($\sim 40$ km s$^{-1}$) allows
the selection of regions free of strong absorption lines in the Lyman
alpha forest. A reliable evaluation of the continuum shape is based on
the careful flux calibration and on the large wavelength interval
covered (5000--9300 ~\AA). A best estimate of $\tau _{_{GP}}\leq
0.02\pm 0.03$ has been found for the Gunn-Peterson optical depth at
the highest absorption redshift observed at this resolution,
$z \simeq 4.3$. The derived baryon density of the
intergalactic medium is $\Omega _{IGM}\mincir 0.01$ if the observed
quasars are the major contributor to the ionizing UV background flux.
This limit, when compared with the total baryon density deduced from
the nucleosynthesis, could imply that most of the baryons are already
in bound systems at $z\sim 5$.

\bigskip
\noindent
{\it Subject headings}: cosmology: observations -- intergalactic medium
-- quasars: absorption lines

\bigskip\bigskip
\centerline{\bf 1. INTRODUCTION}
\bigskip

The absence of the H Ly$\alpha$ absorption trough in the observed spectra of
high redshift quasars (Gunn-Peterson effect) implies that any smoothly
distributed intergalactic medium (IGM) present along the line of sight to the
quasars must be highly ionized at early epoch (Gunn \& Peterson 1965).

In fact, any estimate of the average depression of the quasar continuum
level shortward of the Ly$\alpha$ emission is made difficult by a crowd of
strong, narrow absorption lines interpreted as Ly$\alpha$ absorptions due
to intervening clouds along the line of sight (Lynds 1971, Sargent et
al. 1980).

Previous estimates of the Gunn-Peterson effect were obtained by
subtracting the line contribution to the measured
average depression, $D_A$, present in low resolution spectra
just shortward of the Ly$\alpha$ emission line (Steidel \& Sargent 1987;
Schneider, Schmidt \& Gunn 1989, 1991;  Giallongo \& Cristiani 1990; Jenkins
\& Ostriker 1991). However, different kinds of biases can affect such indirect
estimates, as pointed out by Giallongo \& Cristiani (1990).
The measured average depression depends on the resolution and spectral
range used to estimate the slope and the level of the continuum longward
of the Ly$\alpha$ emission. The estimate of the line
contribution  to the absorption suffers from the poor knowledge
of the line statistics, exspecially at high redshift.

Recently, it has become feasible to obtain at 4-m class
telescopes high resolution ($R > 20000$) data of relatively faint
quasars which are well calibrated in flux and extended over a wide wavelength
interval. This leads to relatively accurate estimates for the continuum
and for the contribution of discrete absorption clouds. Moreover, the
strength of the Gunn-Peterson effect increases rapidly
with redshift even assuming a constant ionizing
background flux (Miralda-Escud\'e \& Ostriker 1990).
Thus, tighter upper limits on the optical depth of the neutral IGM
or on the number of ionizing sources
can be placed by observations of faint QSOs at the highest redshifts.

Webb et al. (1992) give an estimate of $\tau \simeq 0.04$ at an average
absorption redshift $z\simeq 3.8$ depending on the assumptions about
the shape and the low cutoff of the line column density distribution.
A more stringent and direct upper limit to the Gunn-Peterson effect has been
given by Giallongo, Cristiani \& Trevese (1992) who found $\tau \simeq
0.01$ at $\langle z \rangle =3$ just measuring in the Ly$\alpha$
forest of PKS 2126-158
the average depression of the quasar continuum level in regions free
of strong absorption lines.

We present here a direct measure of the GP effect from observations
of a high redshift (z=4.7) quasar and discuss the
relevant consequences for the physical state of the IGM at this
high redshift.

\bigskip\bigskip
\centerline{\bf 2. OBSERVATIONS AND DATA ANALYSIS}
\bigskip

\centerline{\it 2.1 Data acquisition and reduction}

\bigskip
The quasar BR1202-0725 ($z_{em}=4.695$) was discovered by
Hazard, Irwin \& McMahon in the
framework of the APM wide field multicolor (BRI) photographic survey
for bright high redshift ($z>4$) quasars (see e.g. McMahon, Irwin \&
Hazard 1993).
The coordinate of this QSO are given in McMahon et al. (1994).

This quasar has been observed at ESO (La Silla) in 1993 March, with
the NTT telescope and the EMMI instrument in the echelle mode (see
D'Odorico 1990), in the framework of an
 ESO key program devoted to the
study of the absorption  spectra of high $z$ quasars
(D'Odorico et al. 1993). The detector used was a LORAL CCD (2048$\times$2048
pixels$^2$).

Two spectra of 8000 and 6850 s exposures were obtained on March 13 and 14
respectively,
covering the wavelength range $4700-8400$ ~\AA. Two other spectra of
7200 and 8000 s were obtained on March 15, covering the wavelength
range $6000-9500$ ~\AA.

The slit was 1.2 arcseconds wide and 15 arcseconds long.
The seeing was in the range 0.8--1.2
arcsec. All spectra were taken at air masses lower than 1.2 and
with the slit oriented along the average direction of the atmospheric
dispersion to minimize the  wavelength dependence of any slit losses.
 The absolute flux calibration was
carried out by observing two standard stars, HR5501 and CD32EG
(Hamuy et al. 1993). Two exposures with slits of 1.2 and 5
arcseconds were obtained for each star in the two wavelength intervals in
order to check the flux calibration procedure.

The data reduction has been carried out using the standard echelle
package described in the 92NOV edition of the MIDAS software (Banse et
al. 1988). Sky subtraction was carried out by sampling the sky above
and below the QSO spectrum.
The fluxes have been dereddened for galactic extinction according to the Savage
and Mathis (1979) curves and adopting a value E$_{B-V}=0.025$ on the basis
of the Burnstein and Heiles (1982) maps.
The weighted mean of the spectra has been obtained
at the resolution $R=7500$ after rebinning at uniform $\Delta \lambda$
bins close to the original sampling. On average the resolution element was 3
pixels wide.
The signal-to-noise ratio ranges from
6 to 12 per pixel in the interval $6000-9300$ ~\AA.

We have checked the accuracy of the flux calibration, comparing
individual calibrated spectra taken on different nights using
the two different stars. We found
differences up to 5 \% in the average flux levels among the
different spectra but any
$\lambda$ dependent trend in the flux difference was confined to within
1\%  of the average flux.

Using the magnitude at $\lambda =1450$ ~\AA~ rest frame, defined by
$m_{\nu}(1450)=-2.5 \log f_{\nu} -48.60$, with
$f_{\nu}=f_{\nu}[1450(1+z)]$ (Oke \& Gunn 1983) and
applying a correction for the slit loss of 20\% as estimated from the
observations of the standard stars through narrow and wide slits, we
obtain $m_{\nu}(1450)=17.9$.

\bigskip\bigskip
\centerline{\it 2.2 The Gunn-Peterson Test}

We adopted the following general procedure:
first of all, regions longward of the quasar
Ly$\alpha$ emission
were selected for the definition of the continuum level. A power-law
continuum was fitted within these regions and extrapolated in the
Ly$\alpha$ forest. Then, regions in the Ly$\alpha$ forest which are
free of strong absorption lines were selected for the estimate
of the local continuum level. The average Gunn-Peterson optical
depth was simply derived from the ratio of the local continuum level
to the extrapolated one.

Since there are few regions which can be used to estimate the continuum
level, it is important to get a spectral range as wide as possible to
constrain the continuum shape. In particular, the spectrum should
extend beyond the CIV emission, since the local minima selected in the
region between Ly$\alpha$ and C~IV emissions can be spuriously enhanced
by the overlapping wings of the other strong emissions, such as O~I~1302,
C~II~1335 and Si~IV~1400. In our case a good flux calibration was
obtained from  $\lambda \sim 5200$ ~\AA~ up to $\lambda \sim 9300$ (Fig.~1).

We chose a region longward of the C~IV emission
line in the interval $\Delta \lambda = 9150 - 9250$ ~\AA, i.e. at the
minimum between C~IV and the weak He~II~1640 emissions.
A second interval to estimate the continuum level
was taken at the minimum between the Ly$\alpha$ and the O~I emissions
($\Delta \lambda = 7226 - 7261$ ~\AA) which could be still
contaminated by an extended wing
of Ly$\alpha +$ N~V line. In this case the overestimate of the
continuum level results in an overestimate of the Gunn-Peterson optical
depth.

The UV continuum of the quasar was then fitted by a simple power-law
of the type $F_{\lambda}=b (\lambda/7244.)^a$ with $b=(0.108\pm 0.003)
\times 10^{-15}$ erg s$^{-1}$ cm$^{-2}$ ~\AA$^{-1}$
and $a=-1.67\pm 0.08$ (i.e. $\alpha _{\nu}=-0.33$).  The two intervals
used for the fitting procedure and the power law fit are shown in Fig.~2.

Finally, 4 regions free of strong absorption lines in the Ly$\alpha$
forest were selected in the interval
$\Delta \lambda \sim 6100 - 6400$ ~\AA~ (Fig.~3). This way any contamination
by the strong Ly$\alpha$ emission and the highest $z\simeq 4.4$ damped system
present longward of $\lambda \sim 6400$ ~\AA~ was avoided. Similarly,
any contamination by the Ly$\beta$ + O~VI emission blend present shortward
of $\lambda \sim 6100$ ~\AA~ was also excluded.

The ratio $I/I_c$ between the observed continuum level in the selected
regions and the extrapolated one in the Ly$\alpha$ forest was
computed, giving an average optical depth $\tau_{GP}=0.02\pm 0.03$
where the error is due to the noise  in the spectrum and to the
slope uncertainty in the extrapolated continuum. We note however that the
quoted uncertainty can be an underestimate if systematic errors
are present. These could be due to unidentified emission line profiles
in  the regions selected for the estimates of the continuum level
and/or to the non-power-law shape of the continuum.

\bigskip\bigskip
\centerline{\bf 3. DISCUSSION AND CONCLUSIONS}
\bigskip

The value of $\tau_{_{GP}}\simeq 0.02\pm 0.03$ obtained from our spectrum in
the interval $\Delta z = 4.1 - 4.3$ can be compared with the one
obtained by Giallongo et al. (1992) from the spectrum of PKS 2126-158
observed at a resolution of $\simeq 23000$. Using the same procedure,
they derived an upper limit $\tau_{GP}\simeq 0.01\pm 0.03$ at an
average redshift $z=3$.

These two upper limits, covering  the redshift interval $z=3-4.3$,
can be used to constrain the density of the IGM within a consistent
scenario for the thermal history of the IGM and for the cosmological
evolution of the ionizing UV background flux.

For a highly ionized IGM, the Gunn-Peterson optical depth can be expressed
as a function of the IGM density assuming ionization equilibrium where
the heating is dominated by photoionization of the UVB; this yields
\begin{equation}
\tau_{_{GP}}(z)=150~ T^{-0.75}(z) h_{50}^3 \Omega_{IGM}^2 (3+\alpha)
(1+z)^{4.5}
J_{-22}^{-1}(z)
\end{equation}
where $H_0=50 h_{50}$ km s$^{-1}$ Mpc$^{-1}$ is the local Hubble constant,
$J=J_{-22} 10^{-22}$ ergs cm$^{-2}$ s$^{-1}$ Hz$^{-1}$ sr$^{-1}$ is
the ionizing UVB flux
($J\propto (\nu/\nu_{912})^{-\alpha}$) and $T$, $\Omega_{IGM}$ are respectively
the temperature and the baryon density of the intergalactic medium
in units of the cosmological critical density.
A value of $\Omega=1$ has been assumed for the cosmological parameter
throughout the paper.

The thermal history of a photoionized IGM at high $z$ depends on the number and
distribution of ionizing sources and on the effects of the helium reionization
(Miralda-Escud\'e \& Rees 1994; Sciama 1994). In particular, Miralda-Escud\'e
and Rees (1994) have shown that the temperature at a given $z$ depends on the
temperature and redshift immediately after the IGM is wholly ionized, since its
successive thermal evolution is mainly due to adiabatic cooling. They showed
that the temperature at the end of the reionization process can assume values
in the interval 20000 - 50000 K.

If the ionization redshift is as high as $z=9$, the final temperature at
$z\sim 4.5$ is almost independent of its initial value, being $T\sim
10^4$ K. If the ionization of the IGM is just completed at $z=5$ then
its density depends on the initial values assumed for the temperature.
Miralda-Escud\'e \& Rees (1994) compute two thermal evolutions for the ionized
IGM starting at $z=5.7$ with $T=16000$ K and $T=48000$ K. We have used
both curves in equation (1) to estimate the baryon density of the IGM.

The evolution of the ionizing UV flux is the other function appearing in
equation (1) and depends on the kind of sources and on their space density. The
observations suggest that quasars could be the main contributors to the UVB.
However the UVB flux level at $z>4$ depends on poorly known details about the
shape and evolution of the quasar luminosity function (see e.g. Warren, Hewett,
\& Osmer 1994). Madau (1992) and Meiksin \& Madau (1993) have computed
different evolutionary paths for the UVB flux assuming various cosmological
evolutions of the number of quasars and different amounts of UV absorption due
to intervening systems such as strong Ly$\alpha$ lines, Lyman limit and Damped
systems.

Given our low value for the Gunn-Peterson optical depth, we choose to
maximize the ionizing flux provided by known quasar statistics in
order to obtain the maximum value of the density of diffuse hydrogen
compatible with the observation (see eq. 1).

Thus we adopt the model where the comoving space density of quasars
remains constant for $z>3$ together with the low continuum opacity
model for the UVB absorption by intervening Lyman absorption systems.
This last model is supported by recent high resolution Ly$\alpha$
samples where a cutoff or a strong steepening of the column
density distribution of the Ly$\alpha$ lines at $N_{HI}\sim 10^{14.5}$
cm$^{-2}$ has been suggested (Giallongo et al. 1993).
The ionizing UV flux is assumed constant for $z>3$,
taking the value $J_{-22}\sim 3$ with $\alpha =0.7$.

 From Eq.\ 1 the IGM baryon density is $\Omega _{IGM}\mincir 0.008, 0.011$
for the 16000 and 48000 K thermal evolution scenarios, respectively.

Thus, the best estimates on the optical depth derived from the data
constrain the IGM baryon density to be $\Omega _{IGM}\mincir 0.01$ almost
independently of the initial condition for the temperature.

Fall \& Pei (1993) have estimated that the contribution of quasars to the UVB
at $z\sim 3$ can raise up to $J_{-22}\simeq 7$ if dust obscuration affects the
quasar statistics. This value has been found by Giallongo et al. (1993) at
about the same redshift from the analysis of the proximity effect in a high
resolution Ly$\alpha$ sample. However it is not clear if this value can be
extrapolated to $z>4$. If this is the case, the baryon density can raise to
$\Omega _{IGM}\mincir 0.012, 0.017$.

If the UV background is not far from the value predicted on the basis
of the observed quasar counts, then much of the baryon density
$\Omega _b=0.05 h_{50}^{-2}$ derived from the
nucleosynthesis (Walker et al. 1991) remains to be explained.
Indeed, optically thick absorbers observed
along the line of sight to quasars as Lyman limit systems and Damped
systems, which are thought to be connected with protogalaxy haloes
and disks, can contribute to the baryon density for a total amount
$\Omega _{LL+D}=0.011$ (Steidel 1990, Lanzetta et al. 1991).
Since the contribution of luminous matter is of the order of
$\Omega _{Lb}\sim 0.003$ (Persic \& Salucci 1992) then even assuming
the highest estimate for the density of the IGM, we obtain a total
baryon density $\Omega _b=\Omega_{IGM}+\Omega_{LL}+
\Omega_D+\Omega_{Lb}=0.028$, i.e. about half of the nucleosynthesis
value. It is to be noted that compensating this deficit by increasing
$\Omega_{IGM}$ to 0.036, requires a value for the ionizing UVB as high as
$J_{-22}\magcir 30$.

This conclusion leaves room to a scenario where most
of the mass in the high redshift universe can be in the Lyman $\alpha$
clouds. In fact, the total mass of these clouds depends on the ionization
ratio and on their sizes. An appreciable contribution of the order of
hundredth can be obtained assuming
ionization equilibrium with the quasar UVB and
spherical diameters as large as 200 kpc
for clouds with column density $N_{HI}\sim 10^{15}$ cm $^{-2}$.

\bigskip\bigskip
\noindent
{\bf ACKNOWLEDGMENTS}\\
E.G. acknowledges the support of ESO as a visiting astronomer in the period
august-december 1993, while this work was being completed.

\bigskip

\centerline{\bf REFERENCES}

\refer {Banse, K., Ponz, D., Ounnas, Ch., Grosbol, P., Warmels, R.
	1988, in Instrumentation for Ground-Based Optical Astronomy:
	Present and Future, ed. L. B. Robinson (Springer, Berlin,
	Heidelberg, New York) 431}
\refer {Burstein, D., \& Heiles, C. 1982, AJ, 87, 1165}
\refer {D'Odorico, S. 1990, The Messenger, 61, 51}
\refer {D'Odorico, S., Cristiani S., Giallongo E., Molaro P.,
       Savaglio S. \& Trevese D. 1993, 3rd Sesto Workshop, in press}
\refer {Fall, S. M., \& Pei, Y. C. 1993, ApJ 402, 479}
\refer {Giallongo, E., \& Cristiani, S. 1990, MNRAS, 247, 696}
\refer {Giallongo, E., Cristiani, S., Fontana, A., \& Trevese, D.
	1993, ApJ, 416, 137}
\refer {Giallongo, E., Cristiani, S., \& Trevese, D. 1992, ApJ, 398,
	L12}
\refer {Gunn, J. E., \& Peterson, B. A. 1965, ApJ, 142, 1633}
\refer {Hamuy, M., Suntzeff, N. B., Heathcote, S. R., Walker, A. R.,
	Gigoux, P., \& Phillips, M. M. 1993, PASP, submitted}
\refer {Jenkins, E. B., \& Ostriker, J. P. 1991, ApJ, 376, 33}
\refer {Lanzetta, K. M., Wolfe, A. M., Turnshek, D. A., Lu, L.,
	McMahon, R. G., \& Hazard, C. 1991, ApJS, 77, 1}
\refer {Lynds, C. R. 1971, ApJ, 164, L73}
\refer {Madau, P. 1992, ApJ, 389, L1}
\refer {McMahon, R. G., Irwin, M. J., \& Hazard C. 1993, in First
	Light in the Universe Stars or QSO's ?, eds. B. Rocca-Volmerange
	et al. (Editions Frontieres)}
\refer {McMahon, R. G., Omomt, A., Bergeron, J., Kreysa, E. \&
	Haslam, C. G. T., 1994, MNRAS, in press}
\refer {Meiksin, A., \& Madau, P. 1993, ApJ, 412, 34}
\refer {Miralda-Escud\'e, J., \& Ostriker, J. P. 1990, ApJ, 350, 1}
\refer {Miralda-Escud\'e, J., \& Rees, M. J. 1994, MNRAS, 266, 343}
\refer {Oke, J. B., \& Gunn, J. E. 1983, ApJ, 266, 713}
\refer {Persic, M., \& Salucci, P. 1992, MNRAS, 258, 14p}
\refer {Sargent, W. L. W., Young, P. J., Boksenberg, A., \& Tytler, D.
	1980, ApJS, 42, 41}
\refer {Savage, B. D., \& Mathis, J. S. 1979, Ann Rev Astr A, 17, 73}
\refer {Schneider, D. P., Schmidt, M., \& Gunn, J. E. 1989, AJ, 98, 1507}
\refer {Schneider, D. P., Schmidt, M., \& Gunn, J. E. 1991, AJ, 101,
	2004}
\refer {Sciama, D. W. 1994, ApJ in press, SISSA preprint 191/93}
\refer {Steidel, C. C. 1990, ApJS, 74, 37}
\refer {Steidel, C. C., \& Sargent, W. L. W. 1987, ApJ, 318, L11}
\refer {Walker, T. P., Steigman, G., Schramm, D. N., Olive, K. A., \&
	Kang, H. 1991, ApJ, 376, 51}
\refer {Warren, S. J., Hewett, P. C., \& Osmer, P. S. 1994, ApJ, in press}
\refer {Webb, J. K., Barcons, X., Carswell, R. F., \& Parnell, H. C.
	1992, MNRAS, 255, 319}

\bigskip
\centerline{\bf FIGURE CAPTION}

\bigskip
\noindent
Fig.~1. Absolute flux distribution of BR1202-0725 smoothed to FWHM$\sim
6$ ~\AA.  The data have not been corrected for slit losses. They are
estimated at 20\% from the observations of the standard stars. The fitted
power law continuum is also shown. The region between 7600 ~\AA~
and 7680 ~\AA~ has been corrected for atmospheric absorption.

\bigskip
\noindent
Fig.~2. Selected regions  of the spectrum
longward of the Ly$\alpha$ emission used for the continuum fitting.
The fitted continuum is also shown.

\bigskip
\noindent
Fig.~3. The four regions in the Ly$\alpha$ forest where the
Gunn-Peterson optical depth has been measured by comparing the
local continuum level with the extrapolated one (continuous line).
The bars mark the intervals over which the ratio has been measured.
\end{document}